# Radiative and non-radiative effects of a substrate on localized plasmon resonance of particles

Murugesan Venkatapathi* and Abhay Kumar Tiwari

*Supercomputer Education and Research Center, Indian Institute of Science, Bangalore – 560012, India*

* Corresponding author: murugesh@serc.iisc.ernet.in

Experiments have shown strong effects of some substrates on the localized plasmons of metallic nano particles but they are inconclusive on the affecting parameters. Here we have used Discrete Dipole Approximation in conjunction with Sommerfeld integral relations to explain the effect of the substrates as a function of the parameters of incident radiation. The radiative coupling can both quench and enhance the resonance and its dependence on the angle and polarization of incident radiation with respect to the surface is shown. Non-radiative interaction with the substrate enhances the plasmon resonance of the particles and can shift the resonances from their free-space energies significantly. The non-radiative interaction of the substrate is sensitive to the shape of particles and polarization of incident radiation with respect to substrate. Our results show that plasmon resonances in coupled and single particles can be significantly altered from their free-space resonances and are quenched or enhanced by the choice of substrate and polarization of incident radiation.

## I. Introduction

The localized surface plasmon resonances (LSPR) of particles with various geometries and material properties have been studied for more than a decade and their potential applications are ubiquitous[1]. Almost all these devices and experiments involve particles on a substrate but the studies have been mostly based on LSPR of isolated particles and thus the effects of the substrate have not been understood or exploited sufficiently. While methods like Discrete Dipole Approximations (DDA) and other electrodynamic calculations using Mie theory are useful in modeling extinction and scattering by arbitrary shaped nano structures[2], the interaction with a substrate has been difficult to include because separable co-ordinates for analytical solutions do not exist and Fresnel coefficients for near-field radiating sources are not trivial[3]. A sphere on a surface was initially studied using bispherical coordinates in the electrostatic limit[4] and so was the change in polarizability of a spheroid due to contact with a surface[5]. Marginal effects on the optical properties of dielectric nano particles due to a substrate were observed in some experiments[6]. Later significant enhancement of coupling between gold nano particles on indium-tin-oxide substrate was measured[7]. Further experiments of silver nano particles on many substrates show these effects of the substrate are significant but that effective medium approximations like Maxwell-Garnett do not account for these effects completely[8]. With many parameters to vary it is also difficult to experimentally elucidate all the effects of the interaction of single or coupled particles and different surfaces at different angles of incidence. Neither is an experimental baseline possible using isolated particles in suspension when particles are non-spherical. A method of images to approximate plasmon resonance of arbitrary shaped particles on a substrate was recently reported[9]; a method to compute the polarizability of sphere clusters on a substrate was derived earlier[10]. A study under electrostatic approximation shows the effect of polarization on the interaction of substrates with plasmons of sphere clusters on it[11]. All the above mentioned methods under electrostatic or quasi-static approximation do not include the momentum conservation at the surface; hence the dependence of the surface interaction on the angle of incidence can not be included satisfactorily. First principles calculation of plasmons using density functional theory is possible for systems smaller than the ones considered here. A recent such study on two dimensional graphene has shown that plasmons can be significantly quenched by substrates and long range coulomb interactions play a primary role[12]. We used the DDA formulation to discretize the particle into small dipole grains as a single dipole representation of the particle cannot include the non-radiative interaction and the effect of the particle shape. In addition the Sommerfeld integral relations are numerically evaluated for each dipole grain in the particle to compute the surface interaction in the near-field. The dipole grains in the particle thus interact with each other directly and through the surface; the interaction of the incident beam with the dipole grains has two such components as well. Coupling of nano structures/particles on a substrate are required in many applications that exploit plasmon resonances[13-15]. The extinction spectra are shown to be different from that of the single and coupled particles in the absence of the substrate. We have numerically studied the extinction of small gold (Au) spheres and cylinders on smooth surfaces to study the plasmons that are not totally localized in the particle[16, 17]. The role of radiative and non-radiative effects of the substrate on the LSPR of spheres and cylinders are elucidated by comparing these results with a pure radiative model of the surface.

## II. Methods

The particles studied were discretized into an assembly of dipole grains that interact with each other as in the traditional DDA formulations. Using numerical calculations of Sommerfeld integral relation, the field from a radiating dipole over a surface can be decomposed into cylindrical components parallel to the surface and a plane wave perpendicular to the surface[18]. This allows for the computation of the interaction of all dipoles with each other, directly and through the surface as well using corresponding Sommerfeld-Fresnel coefficients (figure 1). The total electric field on a dipole grain *i* in the particle volume can be represented as

$$\overline{E}_{tot,i} = \overline{E}_{o,i} + \overline{E}_{direct,i} + \overline{E}_{reflected,i} \tag{1}$$

The polarizability $\alpha$ of the dipole grain is computed from the refractive index of the material using the Lattice-Dispersion-Relation[19] and equation (1) can be further qualified as

$$(\alpha_i)^{-1}\overline{P}_i - \overline{E}_{direct,i} - \overline{E}_{reflected,i} = \overline{E}_{o,i} \qquad \text{where } \overline{E}_{direct,i} = \frac{k_0^2}{\varepsilon_0}\sum_{j\neq i}\overline{G}_{ij}\cdot\overline{P}_j \tag{2}$$

$$\text{and } \overline{G}_{ij} = \left[\overline{I} + \frac{\nabla\nabla}{k^2}\right] g(R) \text{ where } g(R) = (4\pi R)^{-1}\exp(ikR), R = \left\|\overline{r}_j - \overline{r}_i\right\| \tag{3}$$

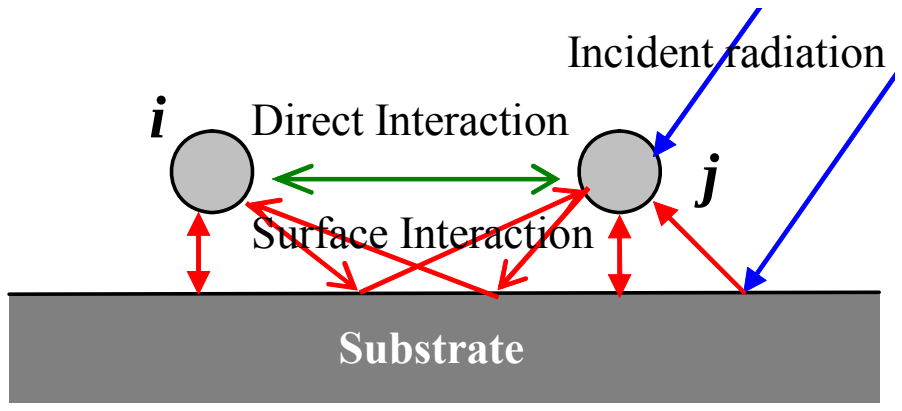

Figure 1: interaction between any two dipole grains $i, j$ in the particle volume (color online)

The determination of the reflected components of the surface on a radiating dipole in the near field involves computing the Sommerfeld relations; for any two dipoles $i$ and $j$ at any distance $z$ from the surface and distance $\rho$ from each other along the surface, the integral in equation (4) has to converge over the complex domain of $k_\rho$ with appropriate branch cuts as the required reflection coefficient $R_{21}$ involves its quadratic roots.

$$\frac{e^{ikr}}{4\pi r} = \frac{i}{4\pi}\int_0^\infty \frac{k_\rho}{k_z} J_o(\rho k_\rho) R_{21} \exp(-ik_z|z|) dk_\rho \tag{4}$$

The reflected component of the electric field due to a radiating dipole in the near field can then be represented as[20]:

$$\bar{E}_{reflected,i} = \sum_{j=1}^{N} (\bar{S}_{ij} + \frac{k_2^2}{\varepsilon_0} \frac{k_1^2 - k_2^2}{k_1^2 + k_2^2} \bar{G}_{ij}^{I}) \cdot \bar{P}_j \qquad (5)$$

Where $\bar{S}_{ij}$ is a 3 x 3 matrix containing Sommerfeld integral terms of the field for dipoles *i* and *j*, and $k_1$ and $k_2$ are the wave numbers for the particle and the surface respectively. The image dyadic Green's function matrix is defined as $\bar{G}_{ij}^{I} = -\bar{G}_{ij} \cdot \bar{I}_R$ where $\bar{I}_R$ is the reflection dyad $\bar{I}_R = \mathrm{e}_x\mathrm{e}_x + \mathrm{e}_y\mathrm{e}_y - \mathrm{e}_z\mathrm{e}_z$.

The Sommerfeld integral relation has no closed analytical solutions and is determined by numerical techniques as shown by Lager and Lytle[21] and later by Mohsen[22]. The computational complexity of above is high and an effective evaluation from tabulated values by interpolation and asymptotic expansion was done as shown by Burke and Miller[23]. The numerical solution to the coupled equation (2) of all the dipole grains in the particle volume ($\bar{P}_i$) involves inverting the matrix containing the polarizability of the dipoles, the Green tensor and Sommerfeld components. This has been accomplished by using a quasi-minimal residual iterative algorithm[24] to obtain the polarization distribution of the volume. The extinction cross sections were found using the approximation applicable to a particle much smaller than the wavelength in dimensions; the sum of the dipole-grain cross sections is the total extinction cross section of the particle.

$$\sigma_{ext} = \Im[\sum_j \bar{E}_{o,j}^{*} \cdot k\bar{P}_j] \qquad (6)$$

**III. Results and Discussion**

The dipole modes of small gold spheres in free space are excited in the range 500-520 nm; a cylindrical particle has additional resonances and more modes have been observed for coupled cylinders with axes perpendicular to each other[15]. We have studied spheres and cylinders (with aspect ratio of 1) of 20nm in diameter and each particle was discretized into a dipole lattice of sufficiently small spacing (2/3 nm) to account for even effects of higher order surface modes of the particle. Any reflection of the incident radiation from a surface directly into the far-field without interaction with the particle is not relevant to this study. The polarizations correspond to the electric field components in the plane of incidence (parallel or *P* polarization) and perpendicular to this plane (*S* polarization); the incident plane is defined by the wave vector of incident radiation and the normal to the substrate. The extinction spectra of gold nano spheres on Indium-Tin-Oxide (ITO) are shown in figure 2. Note that unlike dielectric particles, the particles at plasmon resonance have an extinction cross section that is much larger than their geometric cross section.

**(a)**

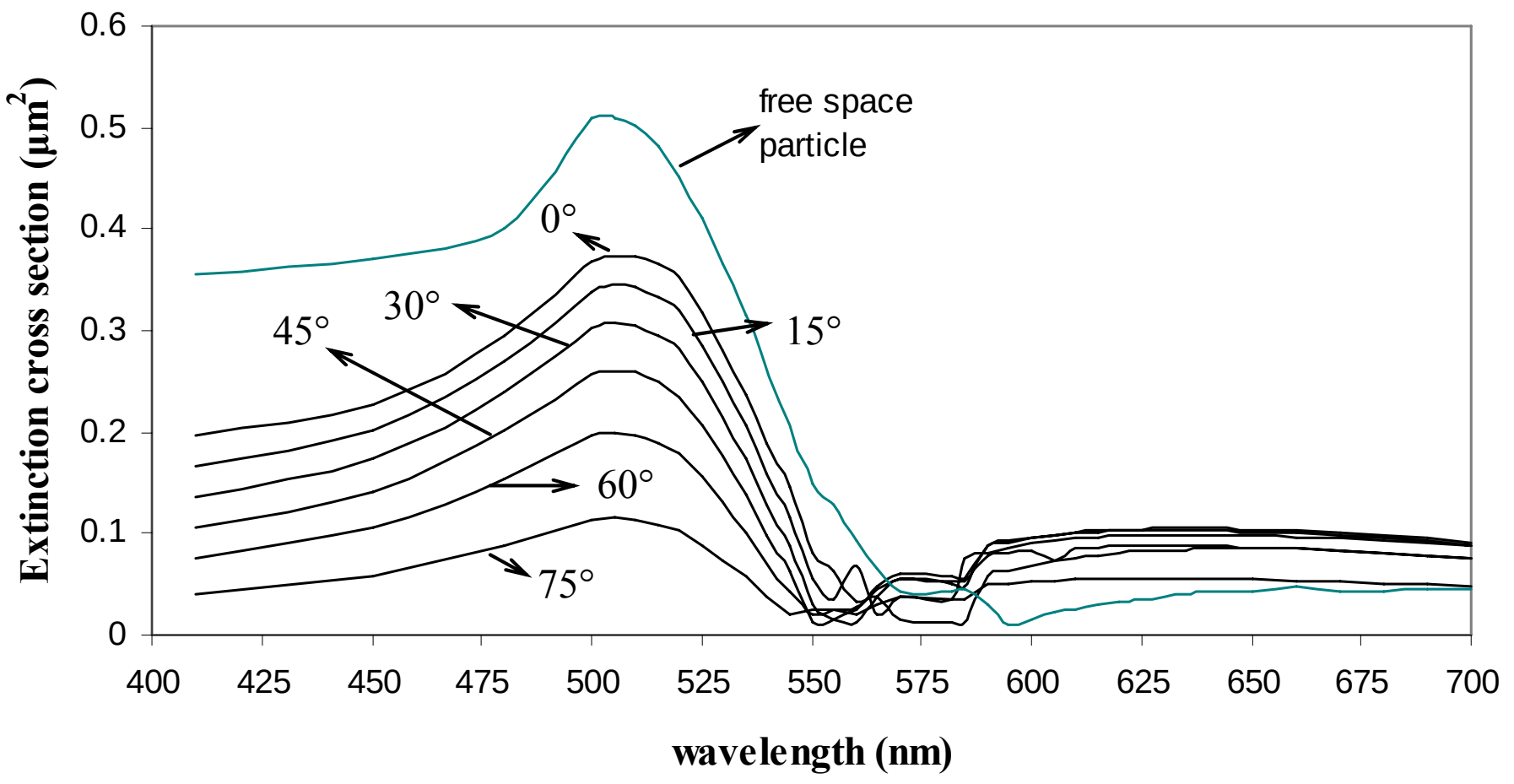

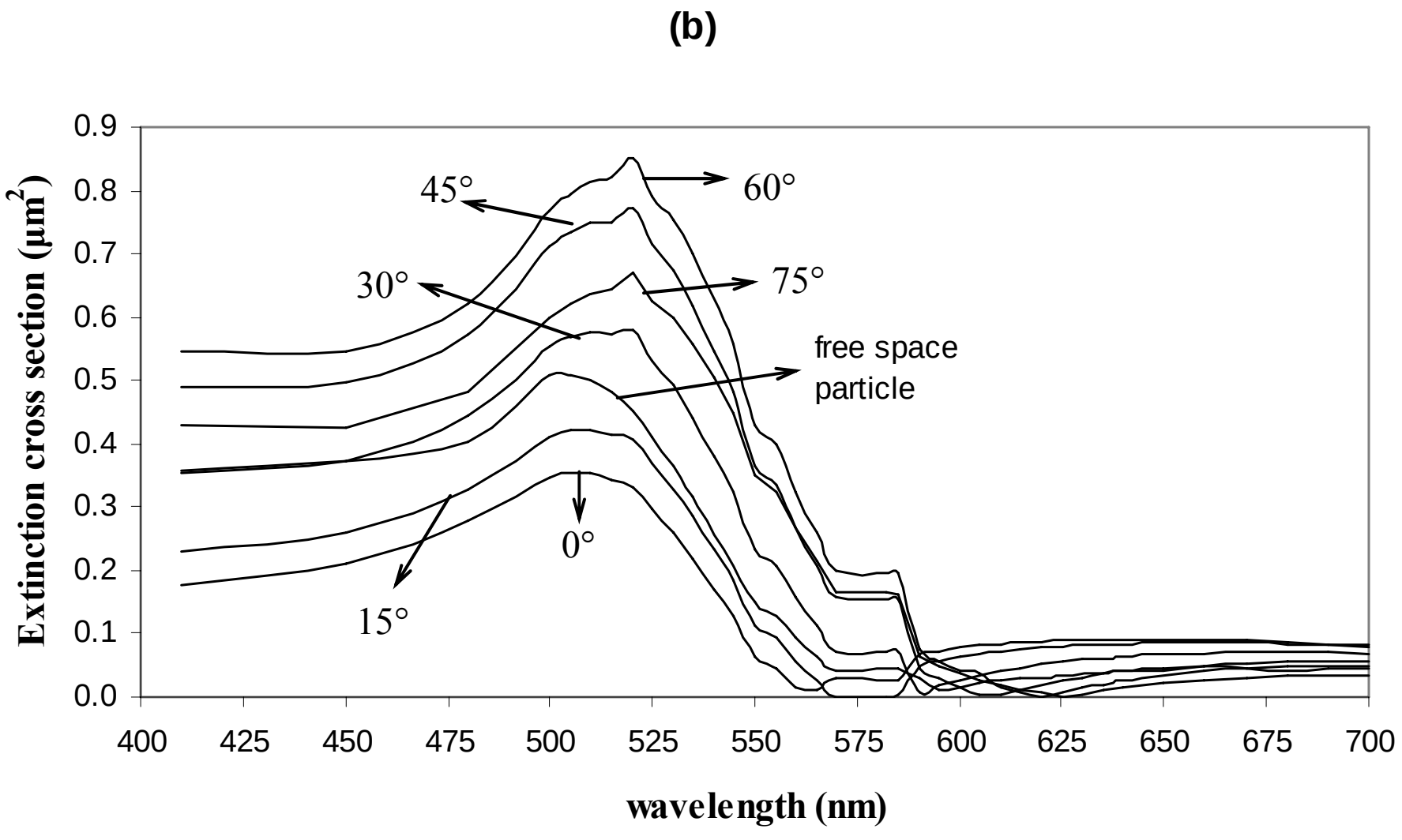

Fig 2: Extinction spectra of 20 nm gold spheres on ITO for different incident angles (with surface normal) a) for *S*-polarized incident radiation b) for *P*-polarized incident radiation

We find the plasmon resonance of spheres quenched for *S*-polarized incident radiation while it is enhanced by *P*-polarized radiation at certain angles due the substrate. To elucidate these results it is illuminating to use a pure radiative model of the surface. This is pertinent for the small particles (<<λ) under LSPR condition where $Q_{abs} >> Q_{sca}$ and the radiative interaction with the surface can be well approximated by the superposition of incident and reflected fields at the surface. The amplitude reflection coefficients '*r'* of a surface given in equation (7) can then be applied to the particle.

$$r_{\parallel} = \frac{n_s \cos\theta_i - n_m \sqrt{(1 - \frac{n_m}{n_s}\sin\theta_i)^2}}{n_s \cos\theta_i + n_m \sqrt{(1 - \frac{n_m}{n_s}\sin\theta_i)^2}} \quad \text{and} \quad r_{\perp} = \frac{n_m \cos\theta_i - n_s \sqrt{(1 - \frac{n_m}{n_s}\sin\theta_i)^2}}{n_m \cos\theta_i + n_s \sqrt{(1 - \frac{n_m}{n_s}\sin\theta_i)^2}} \tag{7}$$

where $n_m$ and $n_s$ are the refractive indices of the surrounding medium and the substrate, $\theta_i$ is the incident angle with the normal to surface.

For a beam reflected from a higher refractive index substrate, note that while the in-plane component (*P* polarization) has a phase shift of π radians at incident angles larger than $\theta_p$ which is called the polarization angle, the perpendicular component (*S* polarization) is always phase shifted by π radians and is hence negative. For the *P*-polarized incident field, one should note that electric field component parallel to the surface has a reflection coefficient of –*r*. Thus, the total electric field on a small particle close to the surface can be approximated as a free space particle irradiated by the modified incident field as in equation (8) for a pure radiative model. The corresponding change in extinction cross section for gold spheres on ITO and silicon at resonance is shown in figure 3.

$$E_\perp = E_{\perp i}\ (1 + r_\perp) \text{ and } E_\parallel = E_{\parallel i}[\cos\theta_i(1 - r_\parallel) + \sin\theta_i(1 + r_\parallel)] \tag{8}$$

The ratio of the extinction cross sections of an isolated particle (at resonance) and that with the surface underneath can be represented by a surface factor. The radiative effect of the surface is strongly dependent on the substrate material and angle of incidence. It is found that silicon can quench the resonance twice as much as ITO for normal incidence (where *S* and *P* polarization are identical). Nevertheless, at larger angle of incidence, silicon can enhance the plasmon resonance significantly by both radiative and non-radiative contributions. The non-radiative effects of the surface can be seen by the difference in the surface factors of the radiative model and the Sommerfeld-DDA model; the charge distribution ($\nabla \bullet \overline{P}$) in the spheres also clearly elucidates this effect (figure 3c).

(a)

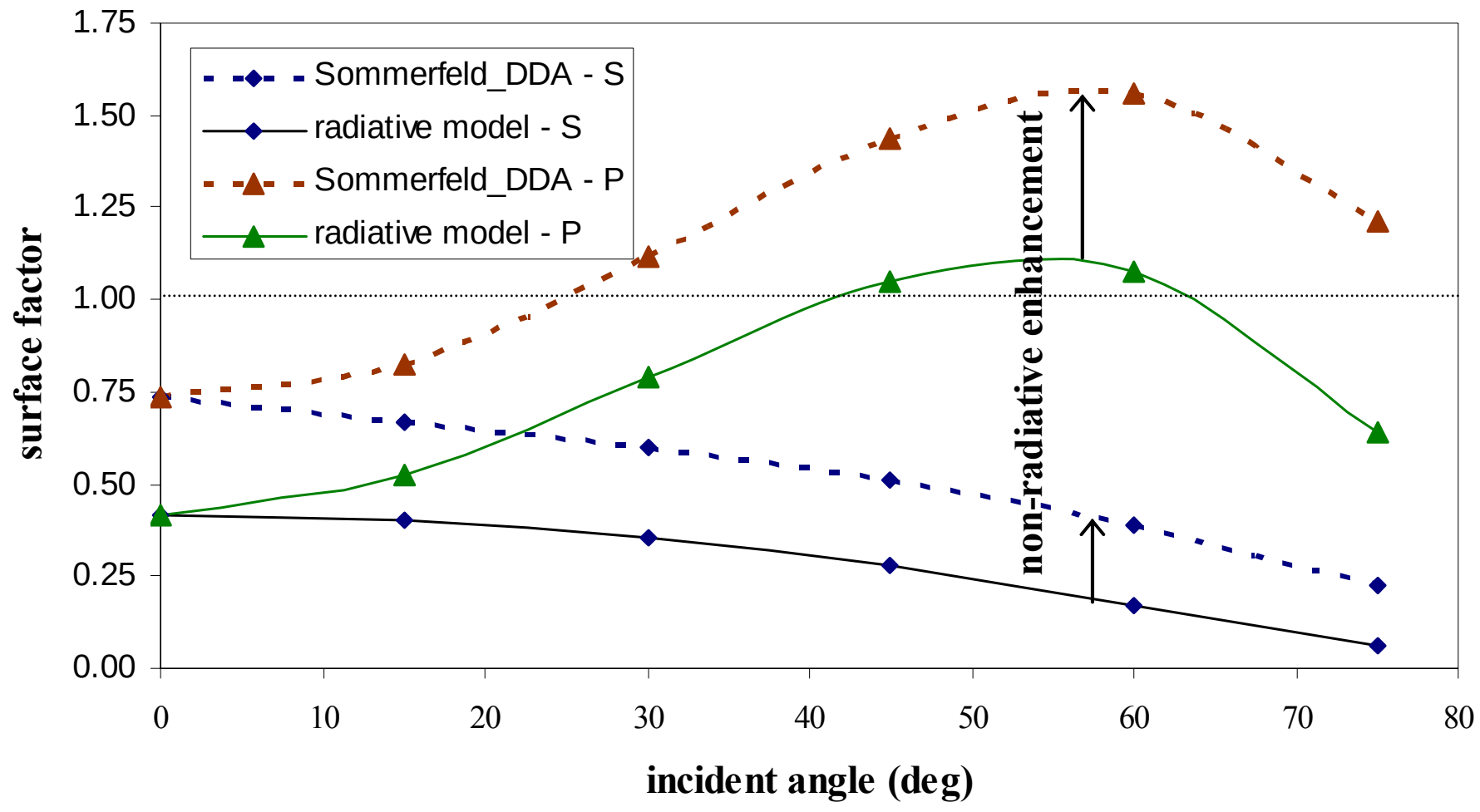

(b)

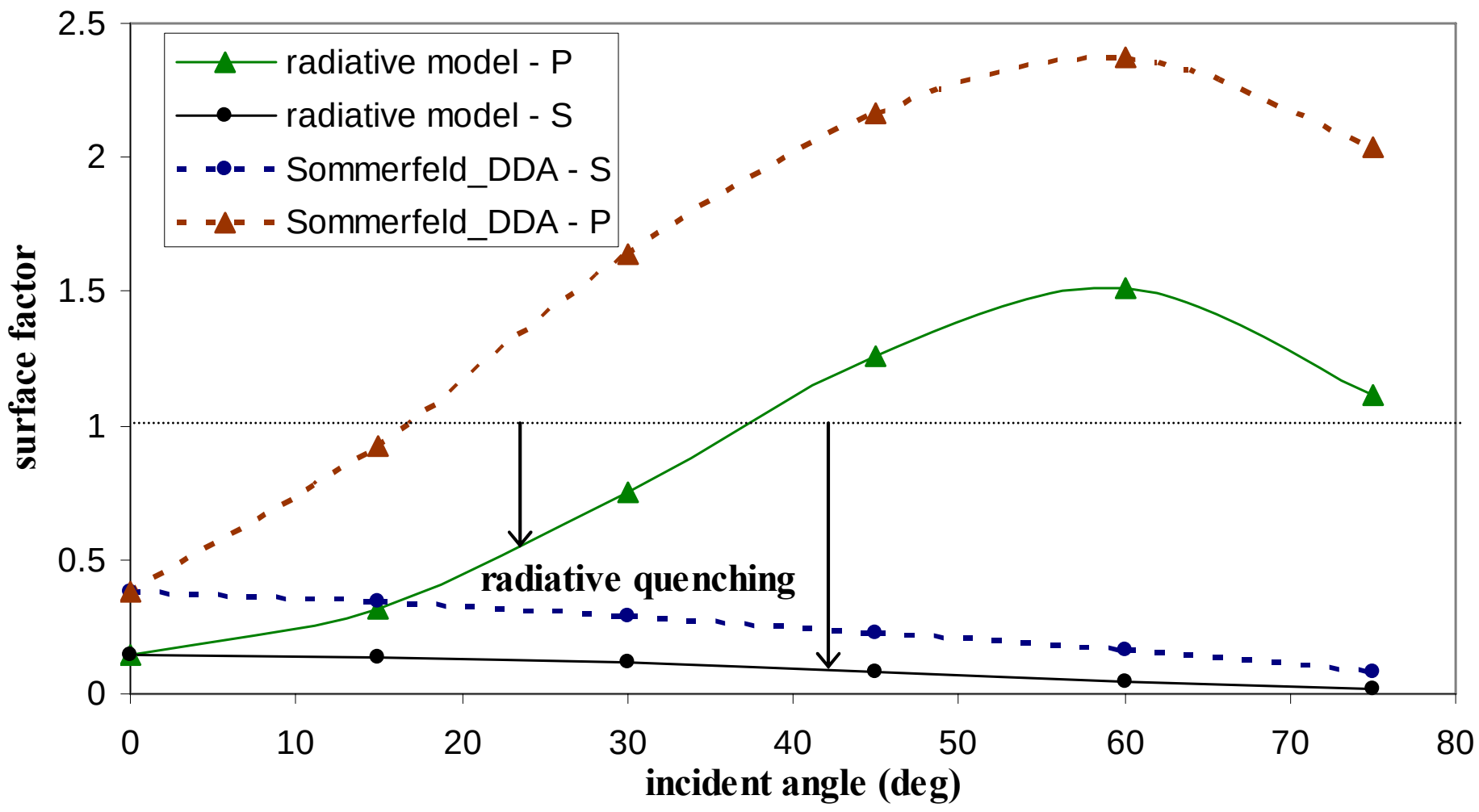

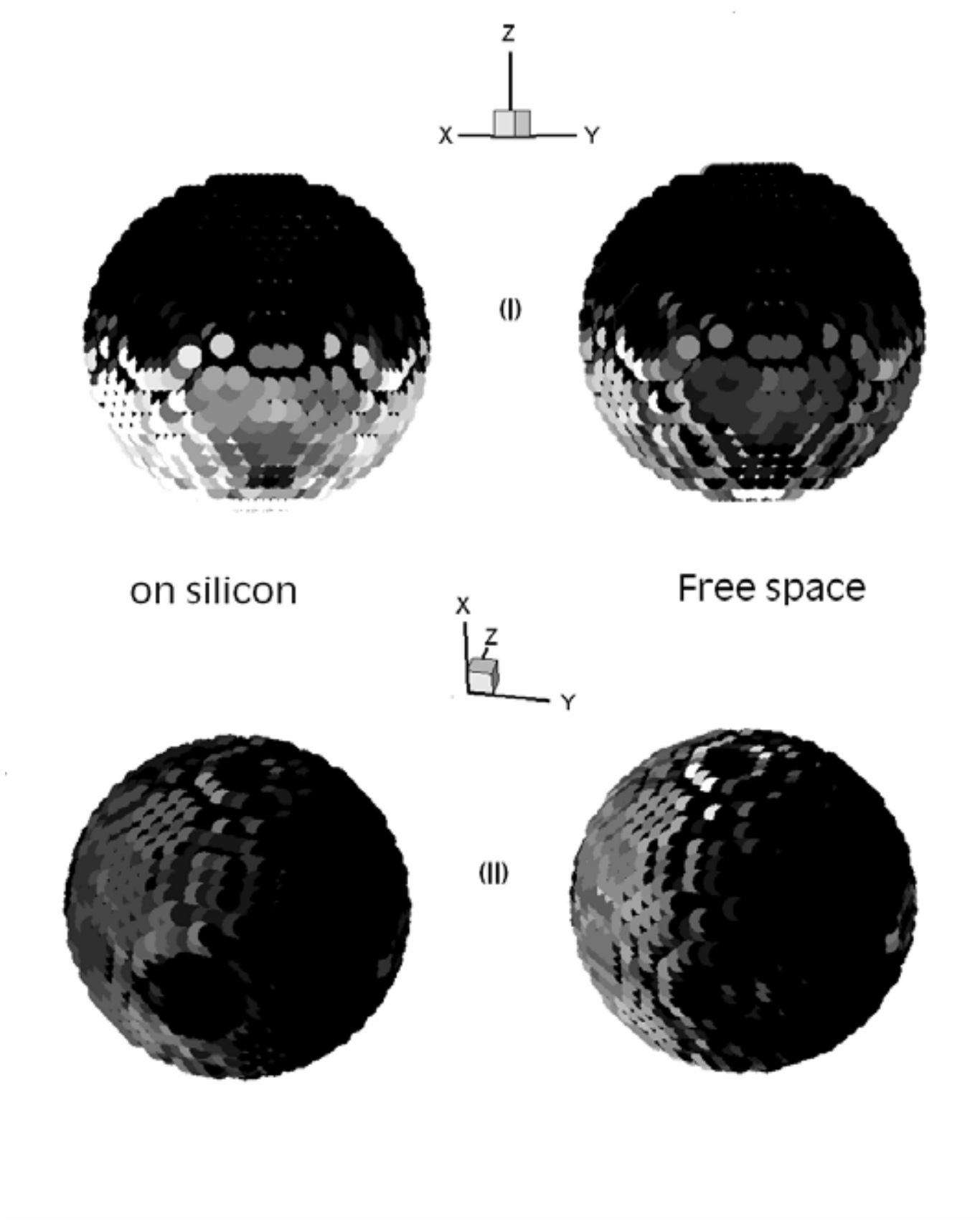

Fig 3:a) surface factor for gold nano-spheres on ITO

b) surface factor for gold nano-spheres on silicon c) relative charge density (-1 to +1) shown in Grayscale (normal to surface is along Z axis): (i) *P*-polarized radiation incident at 60° with –Z axis and (ii) *S*/*P* polarized radiation at normal incidence (electric field vector along Y axis). *Note: above results are for incident radiation of 505nm wavelength*

It is observed that the plasmon resonances are radiatively quenched by a substrate (surface factor <1) for *S*-polarized radiation. In this case, the non-radiative enhancement of the surface is weak as the charge oscillations in the sphere are parallel to the surface. The non-radiative surface enhancement increases with the angle of incidence when the charge oscillation approaches the normal to the surface, as expected for the *P*-polarized radiation. It is evident that the radiative effect/factor of a surface is relatively independent of the shape and size parameters (dimension/wavelength) for particles much smaller than the wavelength. But the non-radiative effect is larger for non-spherical particles and is sensitive to the dispersion properties of the substrate. Thus the non-radiative interaction with the substrate can shift the free-space plasmon resonance; also the resulting non-radiative enhancement can be greater than the radiative quenching of plasmon resonances at near-normal incidence. Numerical calculations of cylinders on ITO substrate show the resonance is indeed shifted to larger wavelengths with respect to the particle in free space, and the resonance is comparable in magnitude to the particle in free space at normal incidence (figure 4a). Such a shift of the free-space plasmon resonance due to the substrate has been observed in experiments on non-spherical particles; silver (Ag) spheroids on vanadium dioxide ($VO_2$) and gold (Au) cylinders on ITO coated glass[8, 9]. It is found here that in the case of gold cylinders on a silicon substrate, the non-radiative enhancement at normal incidence is quite weak compared to radiative quenching, and the plasmon resonance is thus quenched strongly (figure 4b). We realize that the previous experimental observations of 'ITO enhanced' plasmon resonances really should be attributed more to quenching by other substrates at near normal incidence. The substrate can also play a crucial role in coupling of plasmons in two or more particles (distance between particles is on the order of their dimensions) on its surface. Coupling of plasmons in different particles is sensitive to direction of the polarization and the alignment of particles even in the absence of a substrate; evident by a comparison between single cylinder in free-space (in figure 4a) and coupled cylinders in free space for the two different incidence (figure 4b). In such particles the non-radiative coupling of a particle with another particle and its coupling with substrate need not be simply additive (Figure 4b). It should also be mentioned that metal substrates offer both stronger radiative quenching and stronger non-radiative enhancements; a thin metallic film between a dielectric substrate and the particles can enhance coupling[7] and will help reduce the dissipation of a purely metallic substrate. Thus the effects of a substrate in the localized plasmon resonances of single and coupled particles are strong, depending on the substrate and the geometrical and material properties of the particles; a factor that can be exploited for devices and applications with such studies.

(a)

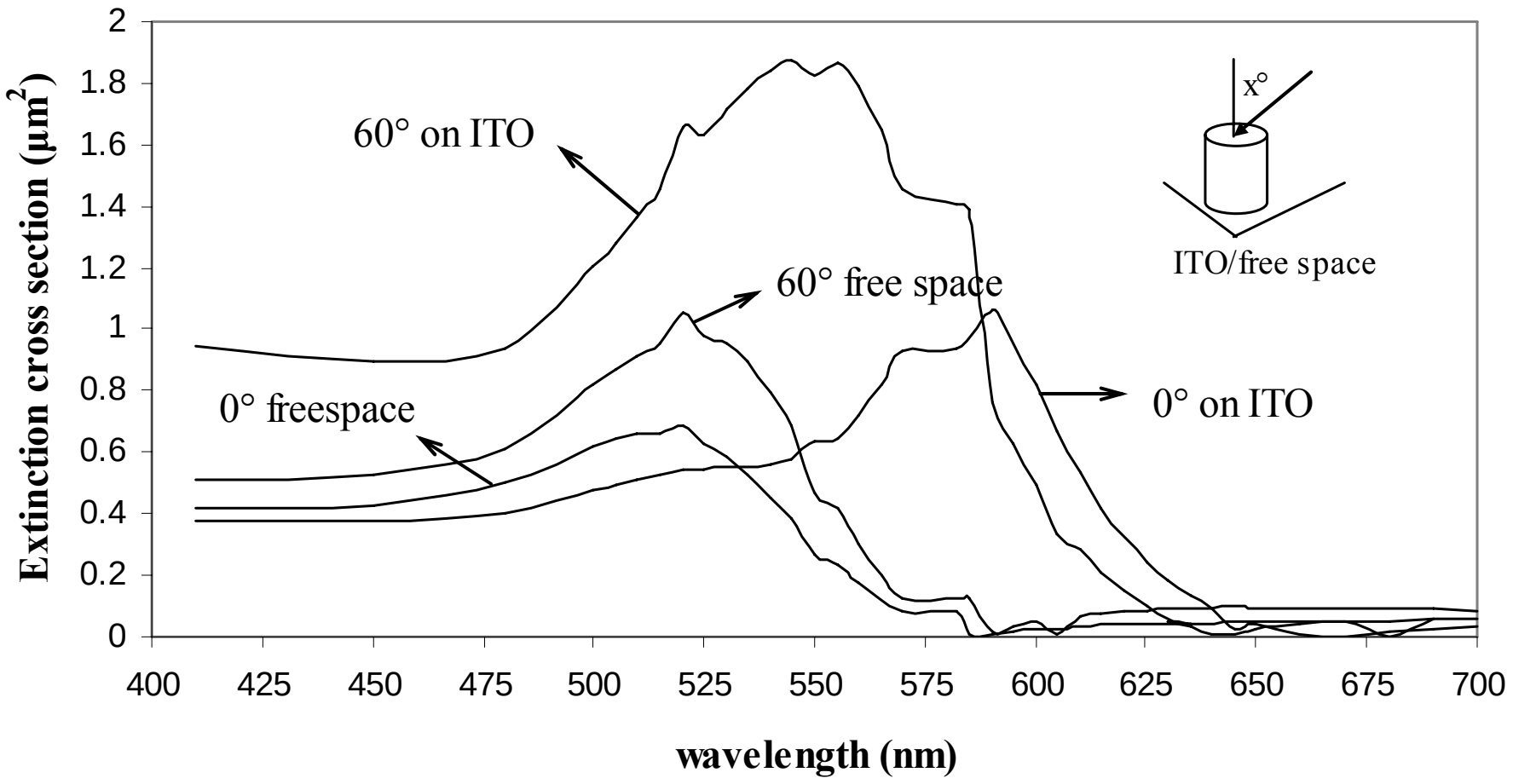

(b)

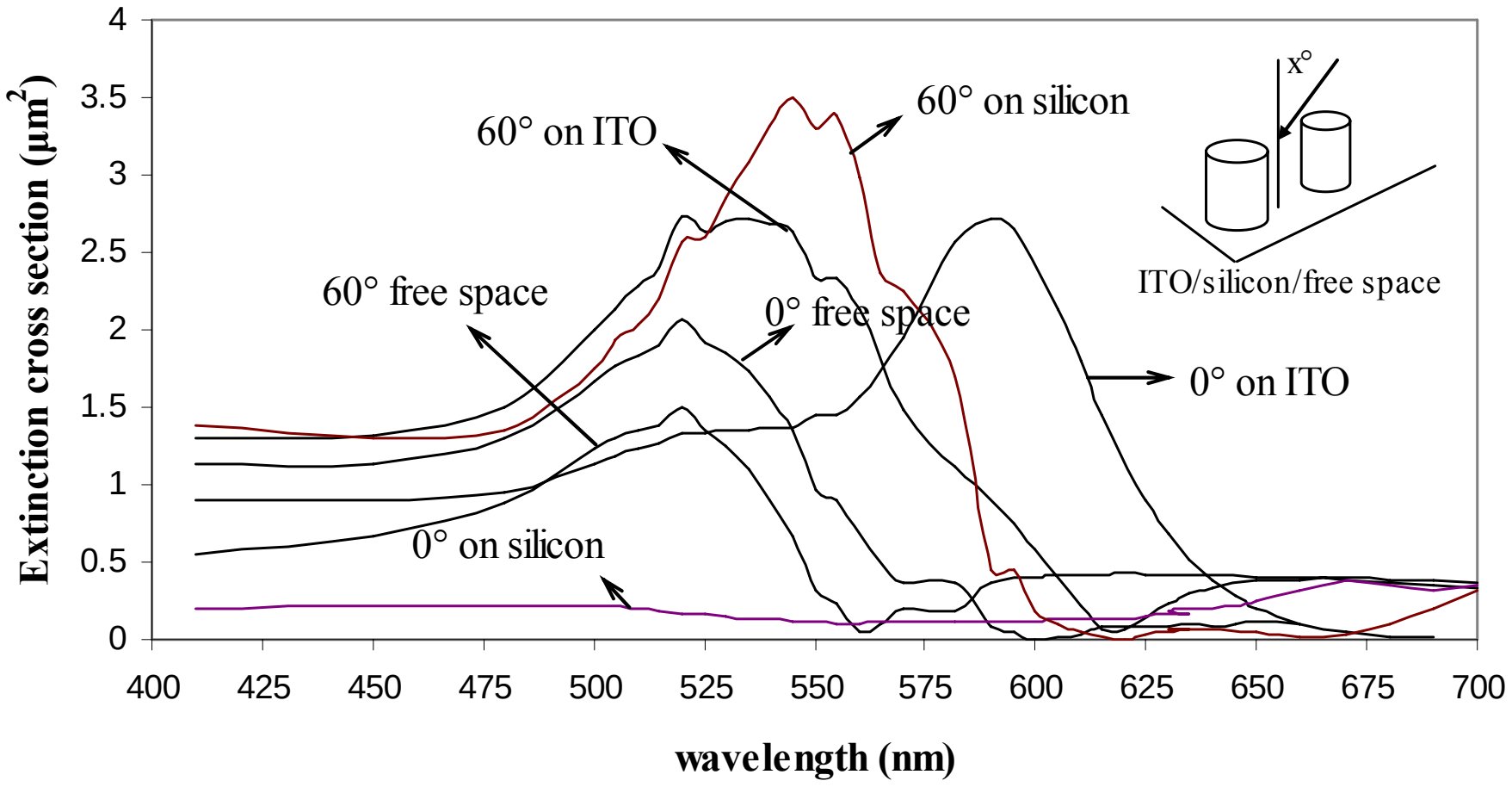

Fig 4: Extinction spectra of 20 nm (length and diameter) gold cylinders for *P*-polarized incident radiation a) isolated cylinder on ITO and free space b) two coupled gold cylinders on silicon and ITO (the centre of cylinders is separated by 30nm in the incident plane)

**Acknowledgements**

We sincerely thank Prof. Dan Hirleman (formerly at Purdue University and currently at University of California, Merced) for sharing with us some of his numerical routines.